\documentclass[12pt]{iopart}

\usepackage{graphicx}
\usepackage{color}
\usepackage{iopams}

\begin{document}

\title{CMB anisotropies at second order III: bispectrum from
products of the first-order perturbations} 

\author{Daisuke Nitta$^1$, Eiichiro Komatsu$^{2,3}$, Nicola Bartolo$^4$, Sabino Matarrese$^4$ and 
Antonio Riotto$^{4,5}$}
\address{$^1$ Astronomical Institute, Tohoku
University, Sendai 980-8578, Japan} 
\address{$^2$ Texas Cosmology Center, the University of Texas at Austin,
Austin, TX 78712, USA}
\address{$^3$ Institute for the Physics and Mathematics of the Universe
(IPMU), University of Tokyo, Chiba 277-8582, Japan}
\address{$^4$ Dipartimento di Fisica ``G. Galilei'', Universita di
Padova, INFN Sezione di Padova, I-35131 Padova, Italy} 

\address{$^5$ CERN, Theory Division, CH-1211 Geneva 23, Switzerland}
\eads{\mailto{nitta@astr.tohoku.ac.jp}, \mailto{komatsu@astro.as.utexas.edu}, 
\mailto{nicola.bartolo@pd.infn.it}, \mailto{sabino.matarrese@pd.infn.it} and \mailto{riotto@mail.cern.ch}}
\date{\today}

\begin{abstract}
We calculate the bispectrum of the Cosmic Microwave Background (CMB)
 temperature anisotropies induced by the second-order fluctuations in the
 Boltzmann equation. In this paper, which is one of a series of papers
 on the numerical calculation of the bispectrum from the second-order
 fluctuations, we consider the terms that are
 products of the first-order perturbations, and leave intrinsically
 second-order terms and perturbations in the recombination history to
 the subsequent papers.  
 We show that the bispectrum has the maximum signal in
 the squeezed triangles, similar to the local-type primordial
 bispectrum, as both types generate
 non-linearities via products of the first-order terms in position space. 
 However, detailed calculations show that their shapes are
 sufficiently different: the
 cross-correlation coefficient reaches $0.5$ at the maximum multipole
 of $l_{max}\sim 200$, and then weakens to $0.3$ at
 $l_{max}\sim 2000$. The differences in shape arise from (i) the way the
 acoustic oscillations affect the bispectrum, and (ii) the second-order
 effects not being scale-invariant. 
 This implies that the contamination of the primordial bispectrum due to
 the second-order effects (from
 the products of the first-order terms) is small. 
 The expected signal-to-noise ratio of the products of the first-order
 terms is $\sim 0.4$ at $l_{max}\sim 2000$ for a full-sky, cosmic
 variance limited experiment. We therefore conclude that the products of the
 first-order terms may be safely ignored in the analysis of the future
 CMB experiments. The expected contamination of the local-form $f_{NL}$
 is $f^{local}_{NL}\sim 0.9$ at $l_{max}\sim 200$, and
 $f^{local}_{NL}\sim 0.5$ at $l_{max}\sim 2000$. 
\end{abstract}

\maketitle

\section{Introduction}
Primordial non-Gaussianity is now recognized as a  powerful probe of the
details of the physics of inflation \cite{review}, as detection of large
primordial non-Gaussianity would rule out all classes of inflation
models that satisfy the following four conditions simultaneously:
single-field, canonical kinetic term, slow-roll, and initially vacuum
state. 

However, the extraction of the primordial non-Gaussianity may not be so
simple, as there are various non-primordial effects that can also generate
non-Gaussianity. Any non-linearities can make initially Gaussian
perturbations non-Gaussian. 

The angular bispectrum, $B_{l_1l_2l_3}$, the harmonic transform of the angular
three-point function, of the Cosmic Microwave Background (CMB) is
often used to measure non-Gaussianity (see, e.g., \cite{komatsuphd}, for
a review). Departures from any of the above conditions (single-field,
canonical kinetic term, slow-roll, and initial vacuum state) would result in
detectable non-Gaussian signals in specific triangle configurations of
the bispectrum. 

When we consider the effects of various non-primordial sources of
non-Gaussianity on the extraction of the primordial signals, we must
specify of which primordial non-Gaussianity we study the contamination
from the non-primordial sources. Multiple-field models, non-canonical
kinetic terms, and initially excited states can produce 
large signals in the squeezed triangles ($l_1\ll l_2\approx l_3$) \cite{lyth/ungarelli/wands:2003},
the equilateral triangles ($l_1= l_2= l_3$) \cite{babich/creminelli/zaldarriaga:2004}, and the
flattened/folded triangles ($l_1\approx l_2\approx l_3/2$)
\cite{chen/etal:2007,holman/tolley:2008}, respectively. 

Throughout this
paper we shall study the contamination of the squeezed triangles,
parametrized in the form of the so-called local form of the bispectrum,
which results from the primordial curvature perturbation (in comoving
gauge) in position space,
$\zeta({\mathbf x})$, given by 
 $\zeta({\mathbf x})=\zeta_L({\mathbf
 x})+\frac35f_{NL}\zeta_L^2({\mathbf x})$, where $\zeta_L$ is a Gaussian
 perturbation, and $f_{NL}$ characterizes the amplitude of the
 local-type non-Gaussianity. Our sign convention is such that the
 temperature anisotropy in the Sachs-Wolfe limit at the first-order in
 perturbations is given by $\Delta T^{(1)}/T=(1/5)\zeta^{(1)}$.
The simplest class of inflation models satisfying all of the four conditions
(single-field,
canonical kinetic term, slow-roll, and initial vacuum state) 
 produce very small non-Gaussian signals: $f_{NL}\sim 10^{-2}$ at the
 horizon crossing during inflation
 \cite{maldacena:2003,acquaviva/etal:2003}, whereas the best limit
 from the WMAP 5-year data with the optimal bispectrum estimator is
 $f_{NL}=38\pm 
 21$ (68\%~CL) \cite{smith}. How much would non-primordial contributions
 account for the measured value of $f_{NL}$?

The CMB bispectrum from the local-type primordial
non-Gaussianity with the linear radiative transfer has been given in
\cite{ks01}, and that arising from non-linearity in gravity has been
considered in \cite{lig06}; however, non-linearities exist also in the
perturbations in the photon-baryon fluid, i.e., non-linearities in the
Boltzmann equation \cite{bar06I,bar06II,pitrou070809}. 

In this paper we calculate the CMB bispectrum, taking into account the
second-order perturbations in 
the Boltzmann equation. We shall include the
second-order terms that are products of the first-order perturbations,
and ignore the intrinsically second-order terms (some of them have been
considered in \cite{bartolo/matarrese/riotto:2006,pitrou,bartolo}), or the
effects of the 
perturbed recombination \cite{khatri09,senatore08I,senatore08II}.
The calculations that also include the intrinsically second-order terms
and the perturbed recombination will be presented
elsewhere.

\section{CMB Bispectrum From Second-order Perturbations}
 \subsection{Definitions}
We expand the temperature fluctuation into the linear (first-order)
part and the second-order part as
\begin{equation}
\frac{\Delta T(\hat{\bf n})}{T}=\frac{\Delta T^{(1)}(\hat{\bf n})}{T}+\frac{\Delta T^{(2)}(\hat{\bf n})}{T}+\ldots.
\end{equation}
The spherical harmonic coefficients of temperature anisotropy, 
$a_{lm}=T^{-1}\int d^2\hat{\bf n}Y^{*}_{lm}(\hat{\bf n})\Delta  T(\hat{\bf n})$,
are therefore expanded as
\begin{equation}
a_{lm}=a_{lm}^{(1)}+a_{lm}^{(2)}+\ldots.
\end{equation}
How do we calculate the second-order part, $a_{lm}^{(2)}$?
This can be calculated by expanding the Boltzmann equation up to the second
order in perturbations \cite{bar06I}.

To expand the Boltzmann equation up to the second order in
perturbations, we first expand the distribution function,
\begin{eqnarray}
f({\bf x},p,{\hat{\bf n}},\eta)=2\left[\exp\bigg\{\frac{p}{T(\eta)e^{\Theta({\bf x},{\hat{\bf n}},\eta)}}\bigg\}-1\right]^{-1},
\end{eqnarray}
up to the second order in perturbations: $\Theta=\Theta^{(1)}+\Theta^{(2)}/2+\ldots,$ and accordingly $f=f^{(0)}+f^{(1)}+f^{(2)}/2+\ldots$.

We compute the fractional perturbation in photon's energy density
at the $i$-th order in perturbations, $\Delta^{(i)}$, by multiplying $f^{(i)}$
by $p$, and integrating over $p^2dp$: 
\begin{equation}
\Delta^{(i)}\equiv \frac{\int dpp^3f^{(i)}}{\int dpp^3f^{(0)}}.
\end{equation}
At the linear order, we recover the usual relation between the
linear fractional temperature fluctuation, $\Theta^{(1)}=\Delta T^{(1)}/T$, and the
linear fractional
energy density perturbation, $\Delta^{(1)}=\delta\rho_\gamma^{(1)}/\rho_\gamma$, i.e.,
$\Delta^{(1)}=4\Theta^{(1)}$. 

At the second order we have 
\begin{eqnarray}
\Delta^{(2)}=4\Theta^{(2)}+16[\Theta^{(1)}]^2,
\end{eqnarray}
which is related to the second-order temperature fluctuation as
\begin{eqnarray}
{\frac{\Delta
T}{T}}^{(2)}&=&\frac{1}{8}\left(\Delta^{(2)}-\langle\Delta^{(2)}\rangle\right)
-\frac{3}{2}\left([\Theta^{(1)}]^2-\langle[\Theta^{(1)}]^2\rangle\right)\nonumber\\
&=&\frac{1}{2}\left(\Theta^{(2)}-\langle\Theta^{(2)}\rangle
+[\Theta^{(1)}]^2-\langle[\Theta^{(1)}]^2\rangle\right),
\end{eqnarray}
where we have subtracted the average of the temperature fluctuation so
that the average of $\Delta T^{(2)}/T$ vanishes. 

We compute $a_{lm}^{(2)}$ from $\Delta T^{(2)}/T$ using 
\begin{eqnarray}
a^{(2)}_{lm}&=&\int d^2\hat{\bf n}Y^{*}_{lm}(\hat{{\bf n}})
{\frac{\Delta T}{T}}^{(2)}\nonumber\\
&=&{\tilde a}^{(2)}_{lm}-\frac{3}{2}\sum_{l'm'}\sum_{l''m''}(-1)^m{\cal G}_{ll'l''}^{-mm'm''}
(a^{(1)}_{l'm'}a^{(1)}_{l''m''}-\langle a^{(1)}_{l'm'}a^{(1)}_{l''m''}\rangle),
\end{eqnarray}
where we define 
\begin{equation}
\fl {\tilde a}^{(2)}_{lm}\equiv\frac{1}{8}\int d^2\hat{\bf n}Y^{*}_{lm}(\hat{\bf n})(\Delta^{(2)}(\hat{\bf n})
-\langle\Delta^{(2)}(\hat{\bf n})\rangle),
\end{equation}
\begin{eqnarray}
\fl {\cal G}_{l_1l_2l_3}^{m_1m_2m_3}&\equiv&\int d^2\hat{\bf n}Y_{l_1m_1}(\hat{\bf n})Y_{l_2m_2}(\hat{\bf n})Y_{l_3m_3}(\hat{\bf n})
\nonumber\\\fl &=&\sqrt{\frac{(2l_1+1)(2l_2+1)(2l_3+1)}{4\pi}}
\left(
\begin{array}{ccc}l_1&l_2&l_3\\0&0&0\\ \end{array}
\right)
\left(
\begin{array}{ccc}l_1&l_2&l_3\\m_1&m_2&m_3\\ \end{array}
\right).
\end{eqnarray}
Here the matrix is the Wigner 3$j$ symbol.

The CMB angular-averaged bispectrum,
$B_{l_1l_2l_3}$, is related to the ensemble average of
$a_{l_1m_1}a_{l_2m_2}a_{l_3m_3}$ as
\begin{eqnarray}
B_{l_1l_2l_3}\equiv\sum_{{\rm all}\,m}
\left(
\begin{array}{ccc}l_1&l_2&l_3\\m_1&m_2&m_3\\ \end{array}
\right)
\langle a_{l_1m_1}a_{l_2m_2}a_{l_3m_3}\rangle.
\end{eqnarray}
This definition guarantees rotational invariance for the bispectrum, and
the Wigner $3j$ symbol ensures that the bispectrum must satisfy triangle
conditions: 
$|l_i-l_j|\le l_k\le l_i+l_j$ for all permutations of indices, and
selection rules: $m_1+m_2+m_3=0$.

The ensemble average is given by
\begin{eqnarray}
\fl\langle a_{l_1m_1}a_{l_2m_2}a_{l_3m_3}\rangle&=&\langle a^{(1)}_{l_1m_1}a^{(1)}_{l_2m_2}a^{(2)}_{l_3m_3}\rangle+cyclic
\nonumber\\
\fl&=&\langle a^{(1)}_{l_1m_1}a^{(1)}_{l_2m_2}{\tilde a}^{(2)}_{l_3m_3}\rangle
-\frac{3}{2}\sum_{l_3'm_3'}\sum_{l_3''m_3''}(-1)^{m_3}{\cal G}_{l_3l_3'l_3''}^{-m_3m_3'm_3''}
\nonumber\\
\fl&&\times
(\langle a^{(1)}_{l_1m_1}a^{(1)}_{l_2m_2}a^{(1)}_{l_3'm_3'}a^{(1)}_{l_3''m_3''}\rangle-
\langle a^{(1)}_{l_1m_1}a^{(1)}_{l_2m_2}\rangle\langle a^{(1)}_{l_3'm_3'}a^{(1)}_{l_3''m_3''}\rangle)
+cyclic ,\label{eq:one}
\end{eqnarray}
where $cyclic$ means that we have to sum the cyclic permutations of
Eq.~(\ref{eq:one}) for indices $(1,2,3) \to(3,1,2) \to(2,3,1)$. 

As we assume that $a_{lm}^{(1)}$'s are Gaussian random variables, 
the four-point function of $a_{lm}^{(1)}$'s in Eq.~(\ref{eq:one}) 
is given by the sum of products of all possible pairs. 
Each pair gives the angular power spectrum, $C_l$:
\begin{eqnarray}
\langle a^{(1)}_{lm}a^{(1)}_{l'm'}\rangle=(-1)^mC_l\delta_{ll'}\delta_{-mm'}.
\end{eqnarray}
We obtain
\begin{eqnarray}
\fl\langle a^{(1)}_{l_1m_1}a^{(1)}_{l_2m_2}a^{(1)}_{l_3'm_3'}a^{(1)}_{l_3''m_3''}\rangle
-\langle a^{(1)}_{l_1m_1}a^{(1)}_{l_2m_2}\rangle\langle a^{(1)}_{l_3'm_3'}a^{(1)}_{l_3''m_3''}\rangle\nonumber\\
=(-1)^{m_1+m_2}C_{l_1}C_{l_2}[\delta_{l_1l_3'}\delta_{-m_1m_3'}\delta_{l_2l_3''}\delta_{-m_2m_3''}
+(1\leftrightarrow 2)]\label{eq:two}.
\end{eqnarray}

Substituting the right hand side of equation (\ref{eq:two}) for the
second term of equation (\ref{eq:one}), and using $l_1+l_2+l_3=$ even,
we obtain the angular averaged bispectrum,  
\begin{eqnarray}
B_{l_1l_2l_3}
=\tilde{B}_{l_1l_2l_3}-3I_{l_1l_2l_3}(C_{l_1}C_{l_2}+cyclic),
\label{eq:16}
\end{eqnarray}
where we have defined the quantities,
\begin{eqnarray}
I_{l_1l_2l_3}\equiv
\sqrt{\frac{(2l_1+1)(2l_2+1)(2l_3+1)}{4\pi}}
\left(
\begin{array}{ccc}l_1&l_2&l_3\\0&0&0\\ \end{array}
\right),\nonumber\\
\end{eqnarray}
and
\begin{eqnarray}
\tilde{B}_{l_1l_2l_3}=\sum_{{\rm all}\,m}
\left(
\begin{array}{ccc}l_1&l_2&l_3\\m_1&m_2&m_3\\ \end{array}
\right)
\langle a^{(1)}_{l_1m_1}a^{(1)}_{l_2m_2}{\tilde a}^{(2)}_{l_3m_3}\rangle
+cyclic.\label{eq:18}
\end{eqnarray}

\subsection{Angular averaged bispectrum from second-order perturbations}

The Boltzmann equation governs the evolution of
$\Delta^{(1)}(k,\mu,\eta)$ and $\Delta^{(2)}({\mathbf k},\hat{\mathbf{n}},\eta)$, 
where $\mu=\hat{k}\cdot\hat{n}$ and $\mathbf{n}$ is the direction of
propagation of photons. Note that for the linear perturbation there is
azimuthal symmetry such that $\Delta^{(1)}$ depends only on the angle
between $\mathbf{k}$ and $\mathbf{n}$; however, for the second-order
perturbation there is no such symmetry.
The Boltzmann equations in Fourier space are given by 
\begin{eqnarray}
{\Delta^{(1)}}'+ik\mu\Delta^{(1)}-\tau'\Delta^{(1)}=S^{(1)}(k,\mu,\eta),\label{eq:1stBol}\\
{\Delta^{(2)}}'+ik\mu\Delta^{(2)}-\tau'\Delta^{(2)}=S^{(2)}(\bf{k,\hat{n}},\eta),\label{eq:2ndBol}
\end{eqnarray}
where the primes denote derivatives with respect to the conformal
 time $\partial/\partial\eta$, 
 $S^{(1)}$ and $S^{(2)}$ are the source functions at the first and
 the second orders, respectively, and $\tau'$ is the differential
 optical depth which is defined by using the mean electron number
 density, $\bar{n}_e$, the Thomson scattering cross-section, $\sigma_T$,
 and the scale factor, $a$, as
\begin{eqnarray}
\tau'=-\bar{n}_e\sigma_Ta. 
\end{eqnarray} 
We expand the angular dependence of $\Delta^{(i)}$ as
\begin{eqnarray}
\Delta_{lm}^{(i)}({\bf k},\eta)=i^{l}\sqrt{\frac{2l+1}{4\pi}}\int
 d^2\hat{\bf n}Y_{lm}^{*}(\hat{\bf n})\Delta^{(i)}({\bf k},\hat{\bf
 n},\eta),
\label{eq:delta_lm}
\end{eqnarray}
and that of the source terms as
\begin{eqnarray}
S_{lm}^{(i)}({\bf k},\eta)=i^{l}\sqrt{\frac{2l+1}{4\pi}}\int d^2\hat{\bf n}Y_{lm}^{*}(\hat{\bf n})S^{(i)}({\bf k},\hat{\bf n},\eta),
\end{eqnarray}
where $i=1,2$.

The source functions relate the observed $a_{lm}$'s to 
the primordial curvature perturbations in comoving gauge,
$\zeta({\mathbf k})$. The relations contain
the linear radiation transfer function, $g_l(k)$, and the second-order
radiation transfer function, $F_{lm}^{l'm'}(k)$, and are given by
\begin{eqnarray}
&&a^{(1)}_{lm}=4\pi(-i)^l\int\frac{d^3k}{(2\pi)^3}g_l(k)Y^*_{lm}(\hat{\bf k})\zeta({\bf k}),
\label{eq:19}\\
&&\tilde{a}^{(2)}_{lm}=\frac{4\pi}{8}(-i)^l\int\frac{d^3k}{(2\pi)^3}\int\frac{d^3k'}{(2\pi)^3}\int d^3k''
\delta^3({\bf k'+k''-k})
\nonumber\\&&\quad\quad\quad\times
\sum_{l'm'}F^{l'm'}_{lm}({\bf k',k'',k})Y^*_{l'm'}(\hat{\bf k})
\zeta({\bf k'})\zeta({\bf k''})\label{eq:20}.
\end{eqnarray}
The linear transfer function is given by
\begin{eqnarray}
\fl g_l(k)=\int_0^{\eta_0} d\eta e^{-\tau}
\left[S_{00}^{(1)}(k,\eta)+S_{10}^{(1)}(k,\eta)\frac{d}{du}+S_{20}^{(1)}(k,\eta)\left(\frac{3}{2}\frac{d^2}{du^2}+\frac{1}{2}\right)\right]j_l(u), 
\end{eqnarray}
where $u\equiv k(\eta_0-\eta)$ and $S_{lm}^{(1)}$ is the standard linear
source function 
(e.g., \cite{cmbfast}):
\begin{eqnarray}
 S_{00}^{(1)}(k,\eta)&=& 4{\Psi^{(1)}}'(k,\eta)-\tau'\Delta_0^{(1)}(k,\eta) ,\label{eq:s00}\\
 S_{10}^{(1)}(k,\eta)&=& 4k\Phi^{(1)}(k,\eta)-4\tau'v_0^{(1)}(k,\eta),\label{eq:s10}\\
 S_{20}^{(1)}(k,\eta)&=& \frac{\tau'}{2}\Delta_2^{(1)}(k,\eta),\label{eq:s20}
\end{eqnarray}
where $\Phi^{(1)}(k,\eta)$ and $\Psi^{(1)}(k,\eta)$ are the metric perturbations at the linear order in the longitudinal gauge:
\begin{eqnarray}
ds^2=a^2(\eta)[-(1+2\Phi^{(1)})d\eta^2+(1-2\Psi^{(1)})\delta_{ij}dx^idx^j] ,\label{eq:1stmetric}\nonumber\\
\end{eqnarray}
and $\Delta_0^{(1)}(k,\eta)$, $\Delta_1^{(1)}(k,\eta)$, and
$\Delta_2^{(1)}(k,\eta)$ are the coefficients of the expansion
 in Legendre polynomials of $\Delta^{(1)}(k,\mu,\eta)$, and $\Delta_l^{(1)}(k,\eta)$
is related to $\Delta_{lm}^{(1)}$ (Eq.~(\ref{eq:delta_lm})) via
$\Delta_{lm}^{(1)}=(-i)^{-l}(2l+1)\Delta_{l}^{(1)}\delta_{m0}$.
The first-order velocity perturbation, $v_0^{(1)}(k,\eta)$, is the
irrotational part of the baryon velocity defined by ${\bf v}({\bf
k})=-iv_0(k)\hat{\bf k}$. 

The new piece, the second-order transfer
function, is the line-of-sight integral of the second-order source terms
in the Boltzmann equation:
\begin{eqnarray}
F^{l'm'}_{lm}({\bf k',k'',k})&=&i^l\sum_{\lambda\mu}(-1)^m(-i)^{\lambda-l'}
{\cal G}_{ll'\lambda}^{-mm'\mu}\sqrt{\frac{4\pi}{2\lambda+1}}
\nonumber\\&\times&
\int_0^{\eta_0}d\eta e^{-\tau}{\cal S}^{(2)}_{\lambda\mu}({\bf k',k'',k},\eta)
j_{l'}[k(\eta-\eta_0)].\nonumber\\
\label{eq:Fdefinition}
\end{eqnarray}
Here, we have introduced a new function, ${\cal S}_{lm}^{(2)}({\bf
 k}',{\mathbf k}'', \mathbf{k},\eta)$, which is defined by the following equation:
\begin{eqnarray}
\fl S_{lm}^{(2)}({\bf k},\eta)=\int\frac{d^3k'}{(2\pi )^3}\int d^3k''\delta^3({\bf k'+k''-k}){\cal S}_{lm}^{(2)}({\bf k',k'', k},\eta)\zeta({\bf k'})\zeta({\bf k''}).
\end{eqnarray}
Basically, ${\cal S}_{lm}^{(2)}({\bf k',k'', k},\eta)$ is the second-order
source function divided by $\zeta({\bf k'})\zeta({\bf  k''})$. 

The explicit expression for $S_{lm}^{(2)}({\bf k},\eta)$ in terms of
perturbation variables is given by Ref.~\cite{bar06I}.
Using equation (\ref{eq:19}) and (\ref{eq:20}), we calculate the first
term in Eq.~(\ref{eq:16}), $\tilde{B}_{l_1l_2l_3}$, as follows:
\begin{eqnarray}
\fl \langle a_{l_1m_1}^{(1)}a_{l_2m_2}^{(1)}\tilde{a}_{l_3m_3}^{(2)}\rangle
=
\frac{(-i)^{l_1+l_2+l_3}}{(2\pi)^3}\sum_{L_3M_3}\prod_i\int d^3k_i\delta^3(\sum_i{\bf k_i})
Y_{l_1m_1}^*({\bf\hat{k}_1})Y_{l_2m_2}^*({\bf\hat{k}_2})Y_{L_3M_3}^*({\bf\hat{k}_3})\nonumber\\
\times g_{l_1}(k_1)g_{l_2}(k_2)P_{\zeta}(k_1)P_{\zeta}(k_2)
\{F_{l_3m_3}^{L_3M_3}({\bf k_1,k_2,k_3})+F_{l_3m_3}^{L_3M_3}({\bf
k_2,k_1,k_3})\}\label{eq:a}, 
\end{eqnarray}
where $P_{\zeta}(k)$ is the power spectrum of $\zeta$ given by the usual definition:
\begin{eqnarray}
\langle\zeta({\bf k_1})\rangle=0,\quad \langle\zeta({\bf k_1})\zeta({\bf k_2})\rangle=(2\pi)^3
\delta^3({\bf k_1+k_2})P_{\zeta}(k_1).
\end{eqnarray}
In order to perform the integral over angles, $\hat{{\bf k}}$, we expand the three-dimensional $\delta$-function using 
Rayleigh's formula,
\begin{eqnarray}
\fl
\delta^3({\bf k_1+k_2+k_3})&=&8\sum_{{\rm all}\,l'm'}i^{l_1'+l_2'+l_3'}{\cal G}_{l_1'l_2'l_3'}^{m_1'm_2'm_3'}
Y_{l_1'm_1'}({\bf\hat{k}_1})Y_{l_2'm_2'}({\bf\hat{k}_2})Y_{l_3'm_3'}({\bf\hat{k}_3})\nonumber\\
\fl &&\times\int drr^2j_{l_1'}(rk_1)j_{l_2'}(rk_2)j_{l_3'}(rk_3),
\end{eqnarray}
and also expand the angular dependence of ${\cal S}_{lm}^{(2)}({\bf
 k_1,k_2,k_3},\eta)$ by introducing the transformed source function,
${\cal S}_{\lambda_1\lambda_2\lambda_3}^{\mu_1\mu_2\mu_3}(k_1,k_2,k_3,\eta)$, as
\begin{eqnarray}
\fl{\cal S}^{(2)}_{\lambda_3\mu_3}({\bf k_1},{\bf k_2},{\bf k_3},\eta)&=&\sum_{\lambda_1,\mu_1}\sum_{\lambda_2,\mu_2}(-i)^{\lambda_1+\lambda_2}
\sqrt{\frac{4\pi}{2\lambda_1+1}}\sqrt{\frac{4\pi}{2\lambda_2+1}}\nonumber\\
\fl&&\times{\cal S}_{\lambda_1\lambda_2\lambda_3}^{\mu_1\mu_2\mu_3}(k_1,k_2,k_3,\eta)
Y_{\lambda_1\mu_1}({\bf\hat{k}_1})Y_{\lambda_2\mu_2}({\bf\hat{k}_2}).\label{eq:28}
\end{eqnarray}
This result shows that ${\cal S}^{(2)}_{\lambda_3\mu_3}({\bf k_1},{\bf
 k_2},{\bf k_3},\eta) = {\cal S}^{(2)}_{\lambda_3\mu_3}({\bf k_1},{\bf
 k_2},k_3,\eta)$, and thus $F_{lm}^{l'm'}({\bf k_1},{\bf k_2},k_3)$
 follows (see Eq.~(\ref{eq:Fdefinition})).

Now we can perform the angular integration of Eq.~(\ref{eq:a}) to obtain
\begin{eqnarray}
\fl\tilde{B}_{l_1l_2l_3}&=&
\frac{4}{\pi^2}(-i)^{l_1+l_2+l_3}\!\sum_{{\rm all}\,m}\sum_{{\rm all}\,l'm'}\sum_{{\rm all}\,\lambda\mu}
\sqrt{\frac{4\pi}{(2\lambda_1+1)(2\lambda_2+1)(2\lambda_3+1)}}i^{l_1'+l_2'+l_3'-\lambda_1-\lambda_2-\lambda_3}\nonumber\\
\fl&&\times\left(
\begin{array}{ccc}l_1&l_2&l_3\\m_1&m_2&m_3\\ \end{array}
\right)
{\cal G}_{l'_1l'_2l'_3}^{m'_1m'_2m'_3}
{\cal G}_{l'_1l_1\lambda_1}^{m'_1-m_1\mu_1}
{\cal G}_{l'_2l_2\lambda_2}^{m'_2-m_2\mu_2}
{\cal G}_{l'_3l_3\lambda_3}^{m'_3-m_3\mu_3}\nonumber\\
\fl&&\times
\prod_{i=1}^{3}\int k_i^2dk_i\int drr^2j_{l_1'}(rk_1)j_{l_2'}(rk_2)j_{l_3'}(rk_3)
g_{l_1}(k_1)g_{l_2}(k_2)P_{\zeta}(k_1)P_{\zeta}(k_2)\nonumber\\
\fl&&\times i^{l_3+l_3'}\int d\eta e^{-\tau}\{ {\cal S}_{\lambda_1\lambda_2\lambda_3}^{\mu_1\mu_2\mu_3}(k_1,k_2,k_3,\eta)+
{\cal
S}_{\lambda_2\lambda_1\lambda_3}^{\mu_2\mu_1\mu_3}(k_2,k_1,k_3,\eta)\}j_{l_3'}[k_3(\eta-\eta_0)]\nonumber\\
\fl&&+cyclic,
\label{eq:tildeBcyclic}
\end{eqnarray}
where we have used the following relation of the Wigner $9j$ symbol,
\begin{eqnarray}
\fl&&(-1)^{l_1'+l_2'+l_3'}\sum_{{\rm all}mm'}
\left(
\begin{array}{ccc}l_1&l_2&l_3\\m_1&m_2&m_3\\ \end{array}
\right)
{\cal G}_{l'_1l'_2l'_3}^{m'_1m'_2m_3'}
{\cal G}_{l'_1l_1\lambda_1}^{m'_1-m_1\mu_1}
{\cal G}_{l'_2l_2\lambda_2}^{m'_2-m_2\mu_2}
{\cal G}_{l'_3l_3\lambda_3}^{m'_3-m_3\mu_3}\nonumber\\
\fl&&\hspace{2cm}=(-1)^{R}I_{l'_1l'_2l'_3}I_{l_1l_1'\lambda_1}I_{l_2l_2'\lambda_2}I_{l_3l_3'\lambda_3}
\Bigg\{
\begin{array}{ccc}l_1&l_2&l_3\\l_1'&l_2'&l_3'\\\lambda_1&\lambda_2&\lambda_3\\ \end{array} 
\Bigg\}
\left(
\begin{array}{ccc}\lambda_1&\lambda_2&\lambda_3\\ \mu_1&\mu_2&\mu_3\\ \end{array}
\right),
\end{eqnarray}
where $R\equiv l_1+l_2+l_3+l_1'+l_2'+l_3'+\lambda_1+\lambda_2+\lambda_3$.
The Wigner 9$j$ symbols have the permutation symmetry:
\begin{eqnarray}
\fl(-1)^{R}
\Bigg\{
\begin{array}{ccc}l_1&l_2&l_3\\l_1'&l_2'&l_3'\\\lambda_1&\lambda_2&\lambda_3\\ \end{array}
\Bigg\}
&=&
\Bigg\{
\begin{array}{ccc}l_2&l_1&l_3\\l_2'&l_1'&l_3'\\\lambda_2&\lambda_1&\lambda_3\\ \end{array}
\Bigg\}
=
\Bigg\{
\begin{array}{ccc}l_1&l_3&l_2\\l_1'&l_3'&l_2'\\\lambda_1&\lambda_3&\lambda_2\\ \end{array}
\Bigg\}
\nonumber\\
\fl&=&
\Bigg\{
\begin{array}{ccc}l_1'&l_2'&l_3'\\l_1&l_2&l_3\\\lambda_1&\lambda_2&\lambda_3\\ \end{array}
\Bigg\}
=
\Bigg\{
\begin{array}{ccc}l_1&l_2&l_3\\\lambda_1&\lambda_2&\lambda_3\\l_1'&l_2'&l_3'\\ 
\end{array}
\Bigg\}\label{eq:9j},
\end{eqnarray}
and the coefficients $I_{l'_1l'_2l'_3}$, $I_{l_1l_1'\lambda_1}$,
$I_{l_2l_2'\lambda_2}$, and $I_{l_3l_3'\lambda_3}$, ensure
$l_1'+l_2'+l_3'={\rm even}$, $l_1+l_1'+\lambda_1={\rm even}$,
$l_2+l_2'+\lambda_2={\rm even}$, and $l_3+l_3'+\lambda_3={\rm even}$,
respectively, which gives $R={\rm even}$. Hence the Wigner $9j$ coefficients
are invariant under the permutations. 

Finally, we obtain the angular averaged bispectrum, 
\begin{eqnarray}
\fl\tilde{B}_{l_1l_2l_3}=\frac{4}{\pi^2}\sum_{{\rm all}\,l'\lambda}
\sqrt{\frac{4\pi}{(2\lambda_1+1)(2\lambda_2+1)(2\lambda_3+1)}}i^{l_3-l_3'+R}
I_{l'_1l'_2l'_3}I_{l_1l_1'\lambda_1}I_{l_2l_2'\lambda_2}I_{l_3l_3'\lambda_3}
\Bigg\{
\begin{array}{ccc}l_1&l_2&l_3\\l_1'&l_2'&l_3'\\\lambda_1&\lambda_2&\lambda_3\\ \end{array} 
\Bigg\}\nonumber\\
\times\int drr^2\prod_{i=1}^{2}\int dk_ik_i^2P_{\zeta}(k_i)g_{l_i}(k_i)j_{l_i'}(rk_i)
\int dk_3k_3^2j_{l_3'}(rk_3)\nonumber\\
\times\int dr' e^{-\tau(r')}j_{l_3'}(r'k_3)
{\cal S}_{\lambda_1\lambda_2\lambda_3}(k_1,k_2,k_3,r')
+perm\label{eq:bis},
\end{eqnarray}
where $r'\equiv \eta_0-\eta$ and we have used the relation of the spherical Bessel function,
$j_l(-x)=(-1)^lj_l(x)$, and have defined the ``angular-averaged source
function,''
\begin{eqnarray}
{\cal S}_{\lambda_1\lambda_2\lambda_3}(k_1,k_2,k_3,r)\equiv\sum_{{\rm all}\mu}
\left(
\begin{array}{ccc}\lambda_1&\lambda_2&\lambda_3\\ \mu_1&\mu_2&\mu_3\\ \end{array}
\right)
{\cal
S}_{\lambda_1\lambda_2\lambda_3}^{\mu_1\mu_2\mu_3}(k_1,k_2,k_3,r)\label{eq:aas}. \end{eqnarray}
Note that $cyclic$ terms in
Eq.~(\ref{eq:tildeBcyclic}) have become $perm$
(permutations) because of invariance of the Wigner $9j$ coefficients under the
permutations. 
 
The final analytic formula (\ref{eq:bis}) we have obtained is a general
formula which can be applied to any second-order perturbations. 
The information about the specific second-order terms is contained in 
the angular-averaged source term, ${\cal S}_{\lambda_1\lambda_2\lambda_3}$
(see Eqs.~(\ref{eq:aas}) and (\ref{eq:28}) for the definition).

For products of the first-order terms, we shall show later that
${\cal S}_{\lambda_1\lambda_2\lambda_3}(k_1,k_2,k_3,\eta)$ does not depend on
$k_3$, i.e., ${\cal S}_{\lambda_1\lambda_2\lambda_3}(k_1,k_2,k_3,\eta)={\cal
S}_{\lambda_1\lambda_2\lambda_3}(k_1,k_2,\eta)$. This property enables us to
integrate Eq.~(\ref{eq:bis}) over $k_3$. We obtain
\begin{eqnarray}
\fl&&\tilde{B}_{l_1l_2l_3}=\frac{2}{\pi}\sum_{{\rm all}\,l'\lambda}
\sqrt{\frac{4\pi}{(2\lambda_1+1)(2\lambda_2+1)(2\lambda_3+1)}}i^{l_3-l_3'+R}I_{l'_1l'_2l'_3}I_{l_1l_1'\lambda_1}I_{l_2l_2'\lambda_2}I_{l_3l_3'\lambda_3}
\Bigg\{
\begin{array}{ccc}l_1&l_2&l_3\\l_1'&l_2'&l_3'\\\lambda_1&\lambda_2&\lambda_3\\ \end{array} 
\Bigg\}\nonumber\\
\fl&&\times\int dr e^{-\tau}\prod_{i=1}^{2}\int dk_ik_i^2P_{\zeta}(k_i)j_{l_i'}(rk_i)g_{l_i}(k_i)
{\cal S}_{\lambda_1\lambda_2\lambda_3}(k_1,k_2,r)+perm,\label{eq:34}
\end{eqnarray}
where $r\equiv \eta_0-\eta$, $R=
l_1+l_2+l_3+l_1'+l_2'+l_3'+\lambda_1+\lambda_2+\lambda_3$, and 
we have used
\begin{eqnarray}
\int dk_3k_3^2j_{l_3'}(rk_3)j_{l_3'}(r'k_3)=\frac{\pi}{2r^2}\delta(r-r').
\end{eqnarray}

Finally, by adding the remaining term in the full bispectrum,
Eq.~(\ref{eq:16}), we obtain
\begin{eqnarray}
\fl&&B_{l_1l_2l_3}=\frac{2}{\pi}\sum_{{\rm all}\,l'\lambda}
\sqrt{\frac{4\pi}{(2\lambda_1+1)(2\lambda_2+1)(2\lambda_3+1)}}i^{l_3-l_3'+R}I_{l'_1l'_2l'_3}I_{l_1l_1'\lambda_1}I_{l_2l_2'\lambda_2}I_{l_3l_3'\lambda_3}
\Bigg\{
\begin{array}{ccc}l_1&l_2&l_3\\l_1'&l_2'&l_3'\\\lambda_1&\lambda_2&\lambda_3\\ \end{array}
\Bigg\}\nonumber\\
\fl&&\times\int dr e^{-\tau}\prod_{i=1}^{2}\int dk_ik_i^2P_{\zeta}(k_i)j_{l_i'}(rk_i)g_{l_i}(k_i)
{\cal S}_{\lambda_1\lambda_2\lambda_3}(k_1,k_2,r)
-\frac{3}{2}I_{l_1l_2l_3}C_{l_1}C_{l_2}+perm.
\label{eq:fullbis}
\end{eqnarray}
The remaining task is to calculate the angular-averaged source term,
${\cal S}_{\lambda_1\lambda_2\lambda_3}(k_1,k_2,\eta)$, which will be given
in the next section.

\section{Second-order bispectrum from products of the first-order terms}

\subsection{Source Term}
The explicit expressions for the second-order source term in Fourier
space are given by Eq.~(5.19) of \cite{bar06I}. 
We will choose the
coordinate system such that $\hat{\bf e}_3=\hat{\bf k}$ in their
expressions, i.e., $\hat{\bf e}_1\perp \hat{\bf k}$, $\hat{\bf e}_2\perp
\hat{\bf k}$, and $\hat{\bf e}_1\perp \hat{\bf e}_2$, 
and adopt the following metric convention:
\begin{eqnarray}
ds^2=a^2(\eta)\left[-e^{2\Phi}d\eta^2+2\omega_idx^id\eta+(e^{-2\Psi}\delta_{ij}+\chi_{ij})dx^idx^j\right],
\end{eqnarray}
where $\Phi=\Phi^{(1)}+\Phi^{(2)}/2$, $\Psi=\Psi^{(1)}+\Psi^{(2)}/2$,
and the shift vector, $\omega_i$, and the transverse and traceless
tensor metric perturbation, $\chi_{ij}$, are already at the second order.
Note that the first-order part of this metric is equivalent  to
Eq.~(\ref{eq:1stmetric}). The second-order source term is
\cite{bar06I,lecture2nd} (also see \cite{pitrou070809})
\footnote{We have corrected the source term given in
Refs.~\cite{bar06I,lecture2nd} for typos, errors, and some missing terms.}
\begin{eqnarray}
\fl S_{lm}({\bf k},\eta)&=&(4{\Psi^{(2)}}'-\tau'\Delta_{00}^{(2)})\delta_{l0}\delta_{m0}+4k\Phi^{(2)}\delta_{l1}\delta_{m0}
-8\omega_{m}'\delta_{l1}-4\tau'v_m^{(2)}\delta_{l1}-\frac{\tau'}{10}\Delta_{lm}^{(2)}\delta_{l2}-4\chi_{m}'\delta_{l2}
\nonumber\\
\fl &+&\int\frac{d^3k_1}{(2\pi)^3}\bigg\{-2\tau'[(\delta_e^{(1)}+\Phi^{(1)})({\bf k_1})\Delta_0^{(1)}({\bf k_2})+2iv_0^{(1)}({\bf k_1})\Delta_1^{(1)}({\bf k_2})]
\delta_{l0}\delta_{m0}
\nonumber\\\fl
&+&4k\Phi^{(1)}({\bf k_1})\Phi^{(1)}({\bf k_2})\delta_{l1}\delta_{m0}
+\tau'[(\delta_e^{(1)}+\Phi^{(1)})({\bf k_1})\Delta_2^{(1)}({\bf k_2})
+2iv_0^{(1)}({\bf k_1})\Delta_1^{(1)}({\bf k_2})]\delta_{l2}\delta_{m0}
\nonumber\\\fl
&+&[8{\Psi^{(1)}}'({\bf k_1})+2\tau'(\delta_e^{(1)}+\Phi^{(1)})({\bf k_1})]\Delta_{l0}^{(1)}({\bf k_2})\delta_{m0}
\bigg\}\nonumber\\
\fl&-&\int\frac{d^3k_1}{(2\pi)^3}{\bf \hat{k}_1}\cdot{\bf \hat{k}_2}
\bigg\{2\tau'v_0^{(1)}({\bf k_1})v_0^{(1)}({\bf k_2})\delta_{l0}-i(-i)^{-l}(2l+1)k_1(\Psi^{(1)}+\Phi^{(1)})({\bf k_1})
\nonumber\\\fl&&\times
\sum_{L}
(2L+1)\Delta_{L}^{(1)}({\bf k_2})\int d\mu P_l(\mu)\frac{\partial P_L(\mu)}{\partial \mu}\bigg\}\delta_{m0}\nonumber\\
\fl&-&2\left[4\Psi^{(1)}\nabla\Phi^{(1)}+4\tau'(\delta_e^{(1)}+\Phi^{(1)}){\bf v}+3\tau'\Delta_0^{(1)}{\bf v}
-\tau'\Delta_2^{(1)}{\bf v}\right]_m\delta_{l1}\nonumber\\
\fl&+&i(-i)^{-l}(-1)^{-m}(2l+1)\sum_{l''}\sum_{m'=-1}^{1}(2l''+1)
\left(
\begin{array}{ccc}l''&1&l\\0&0&0\\ \end{array}
\right)
\left(
\begin{array}{ccc}l''&1&l\\0&m'&-m\\ \end{array}
\right)
\nonumber\\\fl&&\times
\left[8\Delta_{l''}^{(1)}\nabla\Phi^{(1)}+2(\Psi^{(1)}+\Phi^{(1)})\nabla\Delta_{l''}^{(1)}
+2\tau'\Delta_{l''}^{(1)}{\bf v}+5\delta_{l''2}\tau'\Delta_2^{(1)}{\bf v}\right]_{m'}
\nonumber\\
\fl&+&14\tau' (-i)^{-l}(-1)^{-m}(2l+1)\sum_{m',m''=-1}^{1}
\left(
\begin{array}{ccc}1&1&l\\0&0&0\\ \end{array}
\right)
\left(
\begin{array}{ccc}1&1&l\\m'&m''&-m\\ \end{array}
\right)
\nonumber\\\fl&&\times
\frac{4\pi}{3}\int\frac{d^3k_1}{(2\pi)^3}v_0^{(1)}({\bf k_1})v_0^{(1)}({\bf k_2})
Y^{*}_{1m'}({\bf \hat{k}_1})Y^{*}_{1m''}({\bf \hat{k}_2})\nonumber\\
\fl&-&2i(-i)^{-l}\sqrt{\frac{3l+1}{4\pi}}\sum_{m',m''=-1}^{1}\sum_L(2L+1)
\int d^2\hat{\bf n}Y_{1m'}(\hat{\bf n})Y_{1m''}(\hat{\bf n})Y^{*}_{lm}(\hat{\bf n})
\frac{\partial P_L(\mu)}{\partial\mu}\nonumber\\
\fl&&\times \left(\frac{4\pi}{3}\right)^2
\int\frac{d^3k_1}{(2\pi)^3}k_1(\Psi^{(1)}+\Phi^{(1)})({\bf k_1})\Delta_{L}^{(1)}({\bf k_2})
Y^{*}_{1m'}({\bf \hat{k}_1})Y^{*}_{1m''}({\bf \hat{k}_2}),\label{eq:35}
\end{eqnarray}
where ${\bf k=k_1+k_2}$, and $\mu=\hat{\bf n}\cdot\hat{\bf k}$

Here, we have introduced several variables that require explanations. 
The first-order electron number density perturbation is defined by
\begin{eqnarray}
n_e=\bar{n}_e(1+\delta_e^{(1)}).
\end{eqnarray}
The first-order velocity perturbation, ${\bf v}^{(1)}({\bf k})$, 
consists only of the scalar (longitudinal) perturbation:
\begin{eqnarray}
{\bf v}^{(1)}({\bf k})=-iv_0^{(1)}\hat{\bf e}_3.
\end{eqnarray}
The second-order velocity perturbation, ${\bf v}^{(2)}({\bf k})$, 
consists of the scalar perturbation, $v_0^{(2)}$, and the
vector (transverse) perturbation, $v_{m}^{(2)}$: 
\begin{eqnarray}
{\bf v}^{(2)}({\bf k})=-iv_0^{(2)}\hat{\bf e}_3+\sum_{m=\pm 1}v_{m}^{(2)}\frac{\hat{\bf e}_2\mp\hat{\bf e}_1}{\sqrt{2}}.
\end{eqnarray}
The second-order shift vector, ${\bf \omega}({\bf k})$, is
decomposed in a similar way:
\begin{eqnarray}
{\bf \omega}({\bf k})=
\sum_{m=\pm 1}
\omega_{m}\frac{\hat{\bf e}_2\mp \hat{\bf e}_1}{\sqrt{2}}.
\end{eqnarray}
In the gauge choice of \cite{bar06I}, there is no scalar mode in the
shift vector.
For the tensor metric perturbation, 
$\chi_{ij}$, we have 
\begin{eqnarray}
\chi_{ij}=-\sqrt{\frac{3}{8}}\sum_{m=\pm 2}\chi_{m}(\hat{\bf e}_1\pm
 i\hat{\bf e}_2)_i 
(\hat{\bf e}_1\pm i\hat{\bf e}_2)_j.
\end{eqnarray}
The quantities, 
$(f{\bf v})_m$ and $(f\nabla g)_m$, are given by
\begin{eqnarray}
(f{\bf v})_m({\bf
 k})=\sqrt{\frac{4\pi}{3}}\int\frac{d^3k_1}{(2\pi)^3}v_0({\bf
 k_1})f({\bf k}-{\bf k_1})Y_{1m}^{*}({\bf\hat{k}_1}),\nonumber\\
\end{eqnarray}
and
\begin{eqnarray}
(f\nabla g)_m({\bf
 k})=-\sqrt{\frac{4\pi}{3}}\int\frac{d^3k_1}{(2\pi)^3}k_1g({\bf
 k_1})f({\bf k}-{\bf k_1})Y_{1m}^{*}({\bf\hat{k}_1}),\nonumber\\
\end{eqnarray}
respectively.

These perturbation variables of the source term can be split into two
parts; the first line of Eq.~(\ref{eq:35}) contains the variables that
are intrinsically second-order. (The variables have superscripts $(2)$,
and $\omega_m$ and $\chi_m$ are also intrinsically second-order.)
Solving for these terms requires solving the full second-order Boltzmann
equations coupled with the Einstein equations.

The other lines contain the terms that are products of two linear
variables. Evaluation of these terms is much easier than that of the
intrinsically second-order terms, as the first-order variables have
already been calculated using the standard linearized Boltzmann code
such as {\sf CMBFAST} \cite{cmbfast}.

Throughout this paper, we shall evaluate only the products of the
first-order perturbations. The intrinsically second-order perturbations
are equally important, and therefore the final results must also include
those second-order terms. We shall also neglect the contribution from
perturbing the recombination history
\cite{khatri09,senatore08I,senatore08II} for now; we shall
present the full results elsewhere. 

For the products of the first-order perturbations, 
the source terms, ${\cal S}_{\lambda_1\lambda_2\lambda_3}$, are 
non-zero only for the following four cases
(for notational simplicity we shall omit the superscripts (1)):
\begin{eqnarray}
\fl{\cal S}_{000}
&=&4i\tau'v_0(k_1)\Delta_1(k_2)+8\Psi'(k_1)\Delta_0(k_2),\nonumber\\
\fl{\cal S}_{110}
&=&\frac{4}{\sqrt{3}}\big\{-5\tau' v_0(k_1)v_0(k_2)
+2k_1(\Psi+\Phi)(k_1)\sum_{L=odd}(2L+1)\Delta_{L}(k_2)\big\},\nonumber\\
\fl{\cal S}_{101}
&=&2i\sqrt{3}\big\{\tau' v_0(k_1)(4\delta_e+4\Phi+2\Delta_0-\Delta_2)(k_2)\nonumber\\
\fl&&+4k_1\Phi(k_1)(\Delta_0-\Psi)(k_2)+k_1\Delta_0(k_1)(\Psi+\Phi)(k_2)\big\},\nonumber\\
\fl{\cal S}_{112}
&=&2\sqrt{\frac{10}{3}}\big\{7\tau' v_0(k_1)v_0(k_2)
-k_1(\Psi+\Phi)(k_1)\sum_{L=odd}(2L+1)\Delta_{L}(k_2)\big\}.
\end{eqnarray}
From these results we find that ${\cal S}_{\lambda_1\lambda_2\lambda_3}$
does not depend on $k_3$, i.e., 
${\cal S}_{\lambda_1\lambda_2\lambda_3}={\cal
S}_{\lambda_1\lambda_2\lambda_3}(k_1,k_2,r)$. 
Note also that ${\cal S}_{011}(k_1,k_2,r)={\cal S}_{101}(k_2,k_1,r)$.
We have obtained these results by performing the following 
summation over $\mu_1$, $\mu_2$, and $\mu_3$:
\begin{eqnarray}
\fl{\cal S}_{\lambda_1\lambda_2\lambda_3}(k_1,k_2,r)
&=&\sum_{\rm{all}\mu}
\left(
\begin{array}{ccc}\lambda_1&\lambda_2&\lambda_3\\ \mu_1&\mu_2&\mu_3\\ \end{array}
\right)
{\cal S}_{\lambda_1\lambda_2\lambda_3}^{\mu_1\mu_2\mu_3}(k_1,k_2,r)\nonumber\\
\fl&=&i^{\lambda_1+\lambda_2}\sqrt{\frac{2\lambda_1+1}{4\pi}}\sqrt{\frac{2\lambda_2+1}{4\pi}}\sum_{{\rm all}\mu}
\left(
\begin{array}{ccc}\lambda_1&\lambda_2&\lambda_3\\ \mu_1&\mu_2&\mu_3\\ \end{array}
\right)\nonumber\\
\fl&&\times\int d^2\hat{\bf k}_1\int d^2\hat{\bf k}_2
Y_{\lambda_1\mu_1}^{*}(\hat{\bf k}_1)Y_{\lambda_2\mu_2}^{*}(\hat{\bf k}_2)
{\cal S}_{\lambda_3\mu_3}({\bf k_1},{\bf k_2},r),
\nonumber\\ 
\end{eqnarray}
where we have used the inverse relation of Eq.~(\ref{eq:28}).

\subsection{Bispectrum from products of the first-order terms}

Since only four combinations of $\lambda_1$, $\lambda_2$, and
$\lambda_3$ are non-zero, we rewrite the expression for the bispectrum,
Eq.~(\ref{eq:fullbis}), as 
\begin{eqnarray}
\fl B_{l_1l_2l_3}&=&\sum_{\lambda_1\lambda_2\lambda_3}
B_{l_1l_2l_3}^{(\lambda_1,\lambda_2,\lambda_3)}+B_{l_1l_2l_3}^{Cl}
=B_{l_1l_2l_3}^{(0,0,0)}+B_{l_1l_2l_3}^{(1,1,0)}
+2B_{l_1l_2l_3}^{(1,0,1)}+B_{l_1l_2l_3}^{(1,1,2)}+B_{l_1l_2l_3}^{Cl}\label{eq:44},
\end{eqnarray}
where we have used $B_{l_1l_2l_3}^{(0,1,1)}=B_{l_1l_2l_3}^{(1,0,1)}$, 
and defined 
\begin{eqnarray}
B_{l_1l_2l_3}^{Cl}\equiv-3I_{l_1l_2l_3}C_{l_1}C_{l_2}+cyclic,
\end{eqnarray}
and
\begin{eqnarray}
\fl B_{l_1l_2l_3}^{(\lambda_1,\lambda_2,\lambda_3)}&\equiv&\frac{2}{\pi}\sum_{{\rm all}\,l'}
\sqrt{\frac{4\pi}{(2\lambda_1+1)(2\lambda_2+1)(2\lambda_3+1)}}
i^{l_3-l_3'+R}I_{l'_1l'_2l'_3}I_{l_1l_1'\lambda_1}I_{l_2l_2'\lambda_2}I_{l_3l_3'\lambda_3}
\Bigg\{
\begin{array}{ccc}l_1&l_2&l_3\\l_1'&l_2'&l_3'\\\lambda_1&\lambda_2&\lambda_3\\ \end{array}
\Bigg\}\nonumber\\
\fl &&\times\int dr e^{-\tau}\prod_{i=1}^{2}\int dk_ik_i^2P_{\zeta}(k_i)j_{l_i'}(rk_i)g_{l_i}(k_i)
{\cal S}_{\lambda_1\lambda_2\lambda_3}(k_1,k_2,r)+perm.
\label{eq:lambdabis}
\end{eqnarray}

To proceed further, we simplify the expression by introducing the
following notation for the integral over $k$ that appears many times:
\begin{equation}
[x]_{ll'}^{(n)}(r)\equiv\frac{2}{\pi}\int dkk^{2+n}P_{\zeta}(k)j_{l'}(rk)g_{l}(k)x(k,r).\label{eq:66}
\end{equation}
This function corresponds to the existing functions in the literature in
the appropriate limits. For example, for $x(k,r)=\pi/2$, this function
is the same as $\beta_{ll'}^{(n)}(r)$ introduced in \cite{lig06}.
In fact, we find that an order-of-magnitude estimate of
$[x]_{ll'}^{(n)}(r)$ is given by $[x]_{ll'}^{(n)}(r)\sim
2\beta_{ll'}^{(n)}(r)/\pi\times x(k=l'/r,r)$ for a smooth function of
$x(k,r)$. 
As $\beta_{ll'}^{(n)}(r)$ is a sharply peaked function at the decoupling
epoch, $r=r_*$, we find that $[x]_{ll'}^{(n)}(r)$ is also sharply peaked
at $r=r_*$.

With these tools in hand,  we shall calculate 
$B_{l_1l_2l_3}^{(0,0,0)}$,  $B_{l_1l_2l_3}^{(1,1,0)}$, 
$B_{l_1l_2l_3}^{(1,0,1)}$, and $B_{l_1l_2l_3}^{(1,1,2)}$ 
in the following subsections.

\subsubsection{$B_{l_1l_2l_3}^{(0,0,0)}$ and $B_{l_1l_2l_3}^{(1,1,0)}$}

The contributions to the bispectrum from the second-order monopole terms
at the decoupling epoch are $B_{l_1l_2l_3}^{(0,0,0)}$  and
$B_{l_1l_2l_3}^{(1,1,0)}$. For the former the second-order monopole is
created from products of the first-order monopole terms. For the latter
it is created from products of the first-order dipole terms.

First, we calculate $B_{l_1l_2l_3}^{(0,0,0)}$:
\begin{eqnarray}
\fl B_{l_1l_2l_3}^{(0,0,0)}&=&\frac{\pi}{2}\sum_{{\rm all}\,l'}i^{l_3-l_3'+R}\sqrt{4\pi}I_{l_1'l_2'l_3'}I_{l_1l_1'0}I_{l_2l_2'0}I_{l_3l_3'0}\nonumber
\Bigg\{
\begin{array}{ccc}l_1&l_2&l_3\\l_1'&l_2'&l_3'\\0&0&0\\ \end{array} 
\Bigg\}\nonumber\\
\fl&\times&\int dr\bigg\{-4g(r)[v_0]_{l_1l_1}^{(0)}[i\Delta_1]_{l_2l_2}^{(0)}
+8e^{-\tau}[\Psi']_{l_1l_1}^{(0)}[\Delta_0]_{l_2l_2}^{(0)}\bigg\}+perm,\label{eq:000-1}
\end{eqnarray}
where $g(r)$ is visibility function defined by
\begin{eqnarray}
g(r)=-\tau'e^{-\tau},\quad \int_{0}^{\eta_0}dr g(r)=1.
\end{eqnarray}
In the first term of the second line of Eq.~(\ref{eq:000-1}), the
readers might wonder why what-appears-to-be-dipole contributions, $v_0$
and $\Delta_1$, appeared. They should be interpreted as the monopole
contributions, as these contributions here represent the absolute
values of the bulk velocities of the electrons and the photons,
respectively, rather than the dipoles. See  the second term on the
second line of Eq.~(\ref{eq:35}),
$2iv_0^{(1)}\Delta_1^{(1)}\delta_{l0}\delta_{m0}$, which contributes
only to the monopole of the source term, $l=0$.

Eq.~(\ref{eq:000-1}) may be simplified further by using 
\begin{eqnarray}
I_{l_1l_1'0}=(-1)^{l_1}\sqrt{\frac{2l_1+1}{4\pi}}\delta_{l_1l_1'}\label{eq:68},
\end{eqnarray}
and 
\begin{eqnarray}
\Bigg\{
\begin{array}{ccc}l_1&l_2&l_3\\l_1'&l_2'&l_3'\\0&0&0\\ \end{array} 
\Bigg\}
=\frac{\delta_{l_1l_1'}\delta_{l_2l_2'}\delta_{l_3l_3'}}{\sqrt{(2l_1+1)(2l_2+1)(2l_3+1)}}.
\end{eqnarray}
We obtain
\begin{eqnarray}
\fl B_{l_1l_2l_3}^{(0,0,0)}=\frac{1}{2}I_{l_1l_2l_3}\int dr\bigg\{-g(r)[v_0]_{l_1l_1}^{(0)}[i\Delta_1]_{l_2l_2}^{(0)}
+2e^{-\tau}[\Psi']_{l_1l_1}^{(0)}[\Delta_0]_{l_2l_2}^{(0)}\bigg\}+perm\label{eq:000-2}.
\end{eqnarray}

Next, we calculate $B_{l_1l_2l_3}^{(1,1,0)}$:
\begin{eqnarray}
\fl &&B_{l_1l_2l_3}^{(1,1,0)}=-\frac{\pi}{6}\sqrt{2l_3+1}\sum_{{\rm all}\,l'}i^{l_1+l_2+l_1'+l_2'}I_{l_1'l_2'l_3}I_{l_1l_1'1}I_{l_2l_2'1}
\Bigg\{
\begin{array}{ccc}l_1&l_2&l_3\\l_1'&l_2'&l_3\\1&1&0\\ \end{array} 
\Bigg\}
\nonumber\\
\fl&&\times\frac{4}{\sqrt{3}}\int dr\bigg\{5g(r)[v_0]_{l_1l_1'}^{(0)}[v_0]_{l_2l_2'}^{(0)}
+2e^{-\tau}[\Psi+\Phi]_{l_1l_1'}^{(1)}\sum_{L=odd}(2L+1)[i\Delta_{L}]_{l_2l_2'}^{(0)}\bigg\}
+perm.
\end{eqnarray}
We simplify this result further by using 
\begin{eqnarray} 
\Bigg\{
\begin{array}{ccc}l_1&l_2&l_3\\l_1'&l_2'&l_3\\1&1&0\\ \end{array} 
\Bigg\}
=-\frac{(-1)^{l_1'+l_2}}{\sqrt{3(2l_3+1)}}
\bigg\{
\begin{array}{ccc}l_1&l_2&l_3\\l_2'&l_1'&1\\ \end{array} 
\bigg\}.
\end{eqnarray}
Both $l_1'$ and $l_2'$ satisfy the triangular conditions demanded by the
Wigner $6j$ symbols: $l_1-1\le l_1'\le l_1+1$ and $l_2-1\le l_2'\le l_2+1$.
The function $I_{l'_1l'_2l_3}$, which contains the Wigner $3j$ symbols of
$(l'_1,l'_2,l_3; 0, 0, 0)$, requires $l'_1+l'_2+l'_3={\rm even}$. The
other functions, $I_{l_1l_1'1}$ and $I_{l_2l_2'1}$, 
require $l_1+l_1'+1={\rm even}$ and $l_2+l_2'+1={\rm even}$,
respectively. These requirements suggest that one may write
$l_1'-l_1=n_1$ and $l_2'-l_2=n_2$, where $n_1$ and $n_2$ are always odd.
With this result and the above triangular conditions, we find that
$n_1$ and $n_2$ can be either $+1$ or $-1$.
From these results we finally obtain
\begin{eqnarray}
\fl&&B_{l_1l_2l_3}^{(1,1,0)}=\frac{2\pi}{9}\sum_{n_1,n_2=\pm1}i^{n_1-n_2}
I_{l_1'l_2'l_3}I_{l_1l_1'1}I_{l_2l_2'1}
\bigg\{
\begin{array}{ccc}l_1&l_2&l_3\\l_2'&l_1'&1\\ \end{array} 
\bigg\}
\nonumber\\\fl&&
\times\int dr\bigg\{5g(r)[v_0]_{l_1l_1'}^{(0)}[v_0]_{l_2l_2'}^{(0)}
+2e^{-\tau}[\Psi+\Phi]_{l_1l_1'}^{(1)}\sum_{L=odd}(2L+1)[i\Delta_{L}]_{l_2l_2'}^{(0)}\bigg\}
+perm.
\end{eqnarray}

\subsubsection{$B_{l_1l_2l_3}^{(1,0,1)}$}

The contribution to the bispectrum from the second-order dipole terms
at the decoupling epoch is $B_{l_1l_2l_3}^{(1,0,1)}$, 
which  is
created from products of the first-order monopole and dipole terms. 
We obtain
\begin{eqnarray}
\fl B_{l_1l_2l_3}^{(1,0,1)}&=&\frac{\pi}{3}\sum_{n_1,n_3=\pm1}i^{n_1+1}
I_{l_1'l_2l_3'}I_{l_1l_1'1}I_{l_3l_3'1}
\bigg\{
\begin{array}{ccc}l_1&l_3&l_2\\l_3'&l_1'&1\\ \end{array} 
\bigg\}
\nonumber\\
\fl&&\times\int dr\bigg\{-g(r)[v_0]_{l_1l_1'}^{(0)}[4\delta_e+4\Phi+2\Delta_0-\Delta_2]_{l_2l_2}^{(0)}\nonumber\\
\fl&&+4e^{-\tau}[\Phi]_{l_1l_1'}^{(1)}[\Delta_0-\Psi]_{l_2l_2}^{(0)}
+e^{-\tau}[\Delta_0]_{l_1l_1'}^{(1)}[\Psi+\Phi]_{l_2l_2}^{(0)}\bigg\}
+perm,
\end{eqnarray}
where $l_1'=l_1+n_1$ and $l_3'=l_3+n_3$.

\subsubsection{$B_{l_1l_2l_3}^{(1,1,2)}$}

The contribution to the bispectrum from the second-order quadrupole terms
at the decoupling epoch is $B_{l_1l_2l_3}^{(1,1,2)}$, 
which  is
created from products of the first-order dipole terms. 
We obtain
\begin{eqnarray}
\fl B_{l_1l_2l_3}^{(1,1,2)}&=&\frac{2}{3\pi}\sqrt{\frac{4\pi}{5}}(-1)^{l_3}\sum_{{\rm all}l'}
i^{l_1+l_2+l_1'+l_2'}
I_{l'_1l'_2l'_3}I_{l_1l_1'1}I_{l_2l_2'1}I_{l_3l_3'2}
\Bigg\{
\begin{array}{ccc}l_1&l_2&l_3\\l_1'&l_2'&l_3'\\1&1&2\\ \end{array} 
\Bigg\}
\nonumber\\
\fl &&\times2\sqrt{\frac{10}{3}}\int dr\bigg\{-7g(r)[v_0]_{l_1l_1'}^{(0)}[v_0]_{l_2l_2'}^{(0)}
-e^{-\tau}[\Psi+\Phi]_{l_1l_1'}^{(1)}\sum_{L=odd}(2L+1)[i\Delta_{L}]_{l_2l_2'}^{(0)}\bigg\}
\nonumber\\
\fl&&
+perm,
\end{eqnarray}
where $l_1'$, $l_2'$, and $l_3'$ satisfy the triangular conditions:
$l_1-1\le l_1'\le l_1+1$, $l_2-1\le l_2'\le l_2+1$, and $l_3-2\le l_3'\le
l_3+2$, which yields the conditions on $n_1=l_1'-l_1$, $n_2=l_2'-l_2$,
and $n_3=l_3'-l_3$ as
 $-1\le n_1\le1$, $-1\le n_2\le1$, and $-2\le n_3\le 2$.

The Wigner $3j$ symbols in $I_{l'_1l'_2l'_3}$, $I_{l_1l_1'1}$,
$I_{l_2l_2'1}$, and $I_{l_3l_3'2}$ require 
$n_1={\rm odd}$, $n_2={\rm odd}$,  $n_3={\rm even}$, and
$l_1+l_2+l_3={\rm even}$; thus, only $n_1,n_2=\pm1$ and $n_3=\pm2,0$ are
allowed. We finally obtain
\begin{eqnarray}
\fl B_{l_1l_2l_3}^{(1,1,2)}&=&-\frac{8}{9}\sqrt{\frac{6}{\pi}}
\sum_{n_1,n_2=\pm1}\sum_{n_3=\pm2,0}
i^{n_1+n_2}
I_{l'_1l'_2l'_3}I_{l_1l_1'1}I_{l_2l_2'1}I_{l_3l_3'2}
\Bigg\{
\begin{array}{ccc}l_1&l_2&l_3\\l_1'&l_2'&l_3'\\1&1&2\\ \end{array} 
\Bigg\}
\nonumber\\
\fl&\times&\int dr\bigg\{7g(r)[v_0]_{l_1l_1'}^{(0)}[v_0]_{l_2l_2'}^{(0)}
+
e^{-\tau}[\Psi+\Phi]_{l_1l_1'}^{(1)}\sum_{L=odd}(2L+1)[i\Delta_{L}]_{l_2l_2'}^{(0)}\bigg\}
+perm.
\end{eqnarray}

\section{Shape and signal-to-noise of the second-order bispectrum from products of the
 first-order terms}

One of the motivations for calculating the second-order bispectrum is to
see how much the second-order effects in gravity and the photon-baryon fluid
contaminate the extraction of the primordial bispectrum. If, for
example, the predicted shape of the second-order bispectrum is
sufficiently different from that of the primordial bispectrum, then one
would hope that the contamination would be minimal. 
To investigate this, we shall compare the numerical results of the
second-order bispectrum with the so-called ``local'' model of the
primordial bispectrum.

We extract the first-order perturbations from the {\sf CMBFAST} code
\cite{cmbfast}. We use the following cosmological parameters: 
$\Omega_{\Lambda}=0.72,\,\Omega_{m}=0.23,\,\Omega_{b}=0.046,\,h=0.70$,  
and assume a power law spectrum, $P_{\zeta}\propto k^{n-4}$, with
$n=1$. We determine the decoupling time, $\eta_*$, from the peak of
the visibility function. In this model we have $c\eta_0=14.9$ Gpc and
$c\eta_*=288$~Mpc. While the most of the signal is generated in 
the region of the decoupling epoch, in the low-$l$ regime we must also
take into account the late time contribution due to the late integrated
Sachs-Wolfe effect; thus, we integrate over the line-of-sight, $r$, in
the following regions: $c(\eta_0-5\eta_*)<r<c(\eta_0-0.7\eta_*)$ for
$l>100$, and $0<r<c(\eta_0-0.7\eta_*)$ for $l\le 100$. 
The step size  is $\Delta r=0.1\eta_*$ around the decoupling epoch, and
we use the same time steps used by {\sf CMBFAST} after the decoupling
epoch.

The local primordial bispectrum is given by \cite{ks01} 
\begin{eqnarray}
B_{l_1l_2l_3}=2I_{l_1l_2l_3}\int_{0}^{\infty}r^2drb_{l_1}^{L}(r)b_{l_2}^{L}(r)b_{l_3}^{NL}(r)
+cyclic,
\nonumber\\
\end{eqnarray}
where
\begin{eqnarray}
b_{l}^{L}(r)\equiv \frac{2}{\pi}\int_{0}^{\infty}k^2dkP_{\Phi}(k)g^{\rm KS}_{Tl}(k)j_l(kr),\nonumber\\
b_{l}^{NL}(r)\equiv \frac{2}{\pi}\int_{0}^{\infty}k^2dkf_{NL}g^{\rm KS}_{Tl}(k)j_l(kr).
\end{eqnarray}
Note that our linear transfer function, $g_l(k)$, is related to 
that of \cite{ks01}, $g^{\rm KS}_{Tl}(k)$, by $g_l(k)=\frac35g^{\rm KS}_{Tl}(k)$. 

\begin{figure}[t]
\begin{center}
\includegraphics[width=0.495\textwidth]{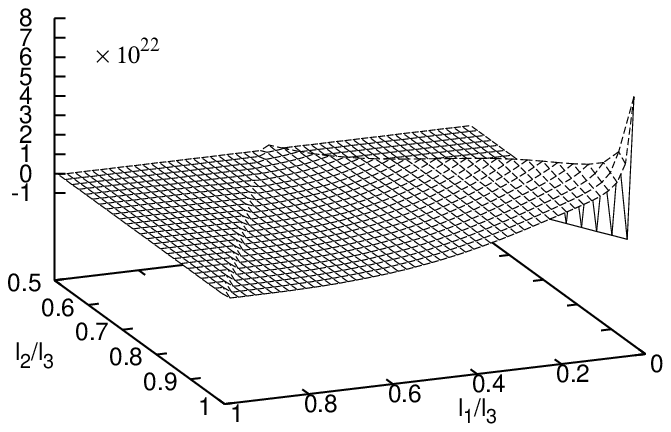}
\includegraphics[width=0.495\textwidth]{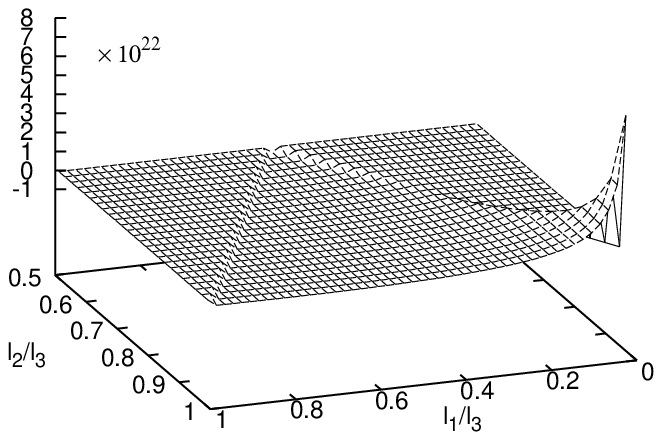}
\end{center}
\caption{
\label{fig:shape1}
Shape dependence of the second-order bispectrum from products of the
 first-order terms (top) and that of the  local primordial bispectrum
 (bottom). We show $l_1l_2\langle
 a_{l_1m_1}^{(1)}a_{l_2m_2}^{(1)}a_{l_3m_3}^{(2)}\rangle
 {({\cal{G}}_{l_1l_2l_3}^{m_1m_2m_3})}^{-1}/(2\pi)^2\times 10^{22}$ as a
 function of $l_1/l_3$ and $l_2/l_3$ where $l_3=200$. Both shapes have
 the largest signals in the squeezed triangles, $l_1\ll l_2\approx
 l_3$. 
}
\end{figure}

Figure~\ref{fig:shape1} shows a shape of the bispectrum generated by the
products of the first-order terms, and compares it to the primordial
bispectrum, for $l_3=200$. 
Both shapes (second-order and primordial) have the largest
signals in the squeezed triangles, $l_1\ll l_2\approx
 l_3$. This is an expected result: the local primordial bispectrum
 arises from the primordial curvature perturbation in position space
 written as $\zeta({\mathbf x})=\zeta_L({\mathbf
 x})+\frac35f_{NL}\zeta_L^2({\mathbf x})$, where $\zeta_L$ is a Gaussian
 perturbation. The second-order bispectrum that we have computed here
arises from the products of the first-order terms, also products in
position space.  
However, these two shapes are slightly different when $l_1/l_3$ is not
so small ($l_1/l_3={\cal O}(0.1)$): the ways in which the radiation transfer
function (which gives the acoustic oscillations) enters into the
bispectrum are different for the products of the first-order terms and
the primordial bispectrum. The primordial bispectrum contains
$j_l(kr_*)g_l(k)$, whereas the second-order bispectrum contains
$j_l(kr_*)g_l(k)x(k,r_*)$ where $x=\Delta_0$, $v_0$, etc., also has the
oscillations. Therefore, the second-order bispectrum has more
interferences between multiple radiation transfer functions. 
Moreover, the second-order effects contain derivatives that the local
primordial effects do not have, which also makes the details of the two
shapes different.

Notice, in particular, that most of these gradients in the source term,
Eq.~(\ref{eq:35}), are contracted with the direction vector, $\hat{\mathbf n}$.
There is only one term that has a scalar product of two wave-vectors,
${\mathbf k}_1\cdot {\mathbf k}_2$, which vanishes in the squeezed limit.
The resulting bispectrum, Eq.~(\ref{eq:44}), resembles that of a local form,
except for the extra powers of $k$ coming from the derivatives. 
These extra powers of $k$ will affect the scale-dependence of the
bispectrum, i.e., the second-order bispectrum is no longer
scale-invariant. Nevertheless,  the largest signal of the bispectrum
still comes from the squeezed configurations, as the number of extra powers of
$k$ from the derivatives in the source term is not large enough to
change the fact that we have the largest contribution when one of $k_1$,
$k_2$, and $k_3$ is very small. In other words, schematically the
bispectrum looks like $B(k_1,k_2,k_3)\sim
(k_1^{m_1}k_2^{m_1})/(k_1^3k_2^3)+cyclic$, where $m_1$ and $m_2$ are the
extra powers of $k$ from the derivatives. Therefore, the largest
contribution is in the squeezed configurations as long as $m_i<3$.

\begin{figure}[t]
\begin{center}
\includegraphics[width=0.495\textwidth]{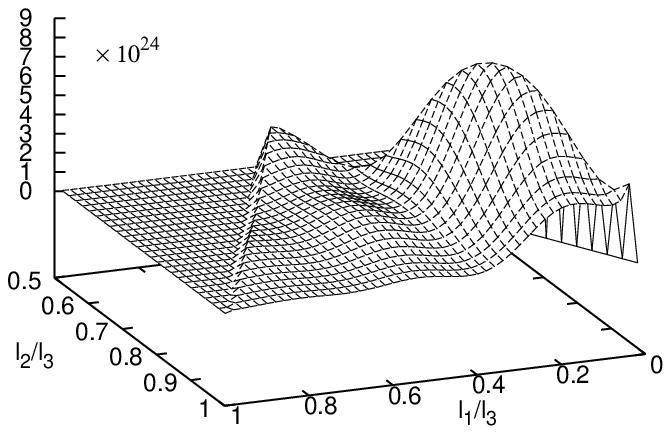}
\includegraphics[width=0.495\textwidth]{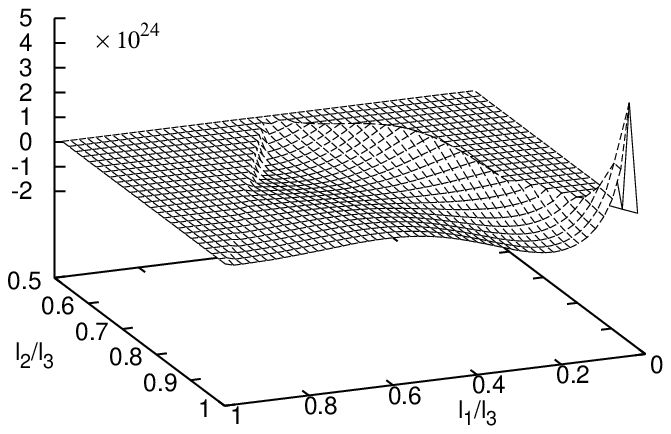}
\end{center}
\caption{
\label{fig:shape2}
Same as Fig.~\ref{fig:shape1} for $l_3=1000$. The acoustic oscillations
 are clearly seen.
}
\end{figure}

Figure~\ref{fig:shape2} shows the same for $l_3=1000$.
The results are similar to those for $l_3=200$, but the acoustic oscillations
are more clearly visible.

How similar are the second-order and the primordial bispectra? What is
the contamination level? We shall quantify the degree to which these
spectra are correlated, as well as the expected signal-to-noise ratio of the
second-order bispectrum, following the standard method given in
\cite{ks01}. Namely, the Fisher matrix for the amplitudes 
of the bispectra, $F_{ij}$, is given by
\begin{eqnarray}
F_{ij}\equiv \sum_{2\le l_1\le l_2\le l_3}\frac{B_{l_1l_2l_3}^{(i)}B_{l_1l_2l_3}^{(j)}}{\sigma_{l_1l_2l_3}^2},
\end{eqnarray}
where 
\begin{eqnarray}
\sigma_{l_1l_2l_3}\equiv 
\langle B_{l_1l_2l_3}^2\rangle-{\langle B_{l_1l_2l_3}\rangle}^2\approx C_{l_1}C_{l_2}C_{l_3}\Delta_{l_1l_2l_3},
\end{eqnarray}
and $\Delta_{l_1l_2l_3}$ takes values 1, 2, and 6 when all l's are different, two of them are equal and all are the same,
respectively. The power spectrum, $C_l$, is the sum of the theoretical
CMB and  the detector noise. Throughout this paper we shall ignore the
noise contribution. In other words, we shall only consider ideal
cosmic-variance limited experiments with full sky coverage.

The signal-to-noise ratio is given by
\begin{eqnarray}
{\left(\frac{S}{N}\right)}_i=\frac{1}{\sqrt{F_{ii}^{-1}}},
\end{eqnarray}
and we define the cross-correlation coefficient between different
shapes $i$ and $j$, $r_{ij}$, as
\begin{eqnarray}
r_{ij}\equiv \frac{F_{ij}}{\sqrt{F_{ii}F_{jj}}}.
\end{eqnarray}

\begin{figure}[t]
\begin{center}
\includegraphics[width=0.6\textwidth]{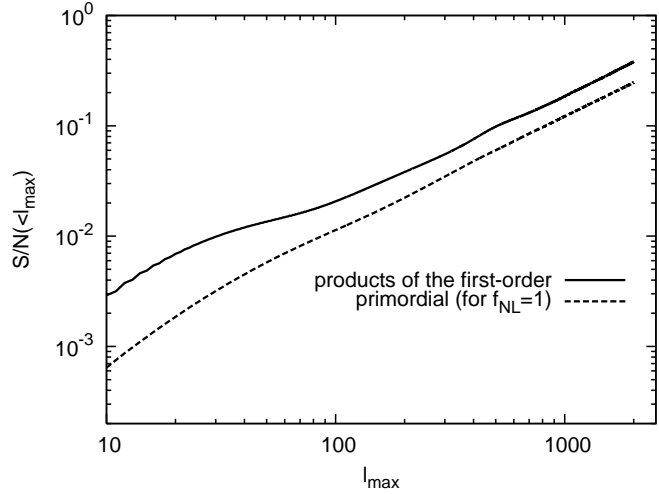}
\end{center}
\caption{\label{fig:SN}
Signal-to-noise ratios for the local primordial bispectrum for $f_{NL}=1$
 (dashed), and the second-order bispectrum from the products of the
 first-order terms (solid), for an ideal full-sky and
 cosmic-variance-limited (noiseless) experiment.
}
\end{figure}

In Fig.~\ref{fig:SN} we show the cumulative signal-to-noise ratio,
summed up to a maximum multipole of $l_{max}$, of the
primordial 
bispectrum, assuming $f_{NL}=1$ and ignoring the second-order
bispectrum, i.e., $(S/N)_{prim}=(F_{prim,prim})^{1/2}$, as well as that of the
second-order bispectrum, ignoring the primordial bispectrum, i.e., 
$(S/N)_{2nd}=(F_{2nd,2nd})^{1/2}$. In both cases $S/N$ increases roughly
as $S/N\propto l_{max}$ (or $\propto \sqrt{N_{pix}}$ where $N_{pix}$ is
the number of independent pixels in the map). A larger contribution to the second-order bispectrum at
$l\lesssim 50$ comes from the terms involving the Integrated Sachs-Wolfe
effect. 
The signal-to-noise ratio of the second-order bispectrum reaches $\sim
0.4$ at $l_{max}=2000$; thus, this signal is undetectable. While our
calculation includes the temperature anisotropy only, 
including polarization would increase the signal-to-noise by a factor of
two at most, which would not be enough to push the signal-to-noise above
unity. 

While the total signal-to-noise does not exceed unity, it may still be
instructive to show which terms of
$B_{l_1l_2l_3}^{(\lambda_1,\lambda_2,\lambda_3)}$ and
$B_{l_1l_2l_3}^{C_l}$ are more important
than the others.  
To do this we show the following quantity:
\begin{eqnarray}
\left(\frac{S}{N}\right)_{ab}\equiv
\left|\sum_{2\le l_1\le l_2\le l_3}\frac{B_{l_1l_2l_3}^{a}B_{l_1l_2l_3}^{b}}
{\sigma_{l_1l_2l_3}^2}\right|^{1/2},
\label{eq:sncomp}
\end{eqnarray}
where $a,b=1$, 2, 3, 4, and 0 correspond to  $(0,0,0)$, $(1,1,0)$,
$(1,0,1)$, $(1,1,2)$, and $C_l$, respectively.

\begin{figure}[t]
\begin{center}
\includegraphics[width=0.6\textwidth]{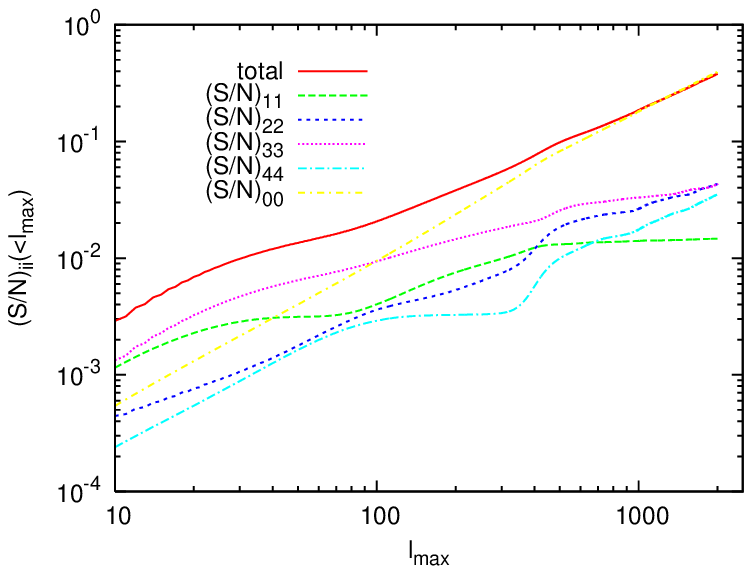}
\includegraphics[width=0.6\textwidth]{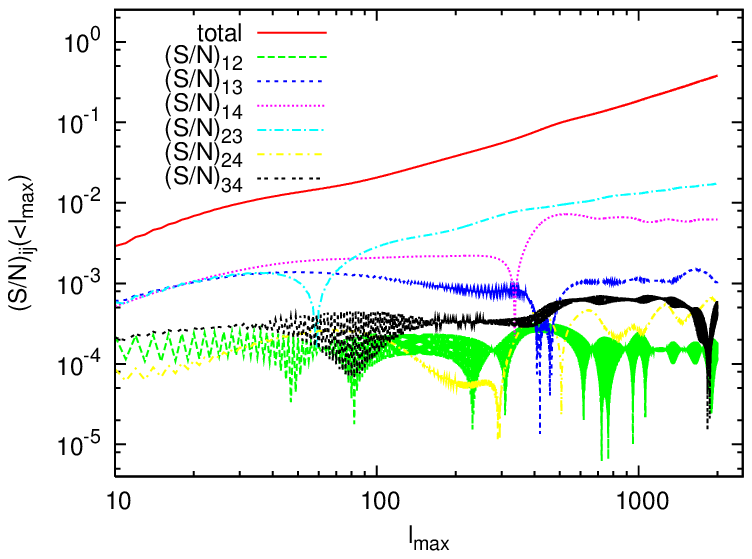}
\includegraphics[width=0.6\textwidth]{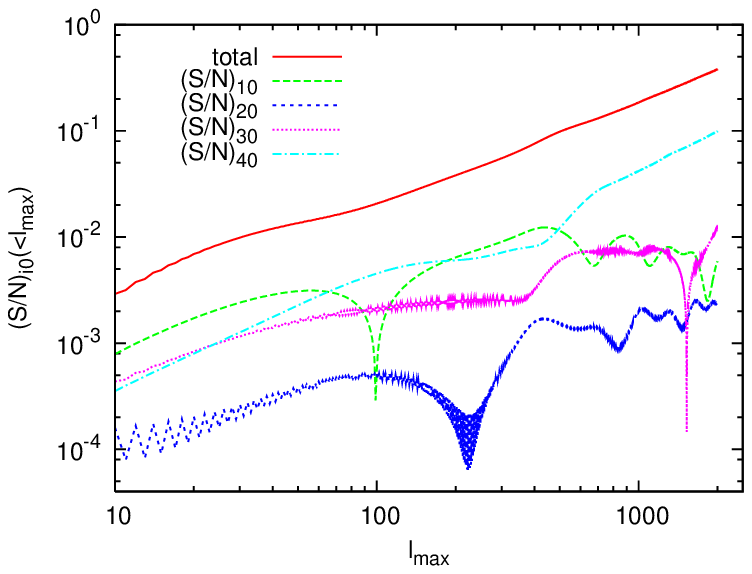}
\end{center}
\caption{\label{fig:SNab}
Absolute values of the contributions to the signal-to-noise ratio from
 each component, $(S/N)_{ab}$, as defined by Eq.~(\ref{eq:sncomp}).
}
\end{figure}

The results are shown in Fig.~\ref{fig:SNab}. We find that 
$(S/N)_{2nd}$ is dominated by
$B_{l_1l_2l_3}^{(\lambda_1,\lambda_2,\lambda_3)}$ for $l\lesssim 100$,
whereas it is 
dominated by $B_{l_1l_2l_3}^{C_l}$ for $l\gtrsim 100$ (see the top panel).

Among $B_{l_1l_2l_3}^{(\lambda_1,\lambda_2,\lambda_3)}$, 
the most dominant term is $(1,0,1)$ (the bispectrum from the
second-order dipole created by the first-order dipole and monopole).
The second most dominant is $(0,0,0)$ (from the second-order monopole
created by the first-order monopole) for $l\lesssim 400$ and $(1,1,0)$
(from the second-order monopole 
created by the first-order dipole) for $l\gtrsim 400$.
The cross terms (middle and bottom panels) are sub-dominant compared to
the auto terms (top panel) at all multipoles.

\begin{figure}[t]
\begin{center}
\includegraphics[width=0.6\textwidth]{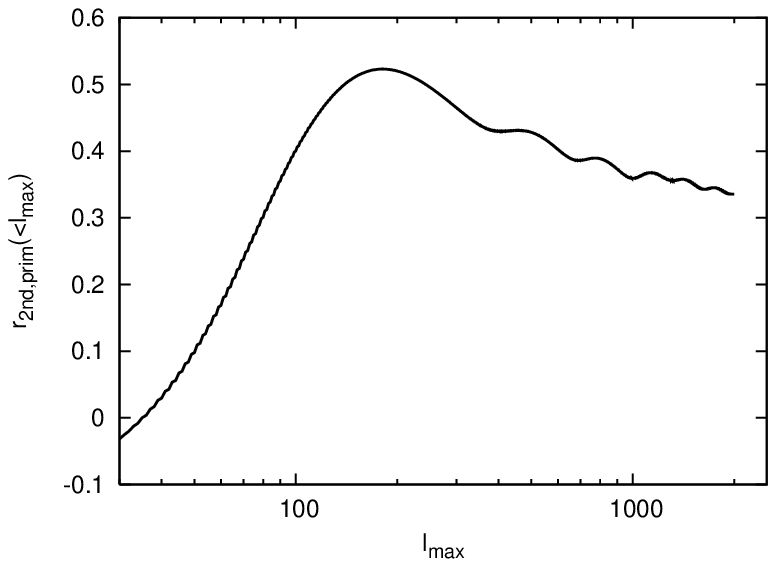}
\end{center}
\caption{\label{fig:r12}
The cross-correlation coefficient between the second-order bispectrum from
 the products of the first-order terms and the local primordial bispectrum. 
}
\end{figure}

How similar are the second-order and the primordial bispectra?
In Fig.~\ref{fig:r12} we show the cross-correlation coefficient between
the second-order bispectrum from  the products of the first-order terms
and the local primordial  bispectrum. The cross-correlation coefficient
reaches $\sim 0.5$ for $l_{max}=200$, and the shapes for $l_3=200$ are 
shown in Fig.~\ref{fig:shape1}. After $l_{max}=200$ the correlation
weakens, and reaches $\sim 0.35$ at $l_{max}=1000$, and 
the shapes for $l_3=1000$ are shown in Fig.~\ref{fig:shape2}.
These results show that the second-order bispectrum from the products of
the first-order perturbations and the local primordial bispectrum are
fairly similar, with a sizable correlation coefficient. The next
question is, ``how large is the contamination of the primordial
bispectrum?''

\begin{figure}[t]
\begin{center}
\includegraphics[width=0.6\textwidth]{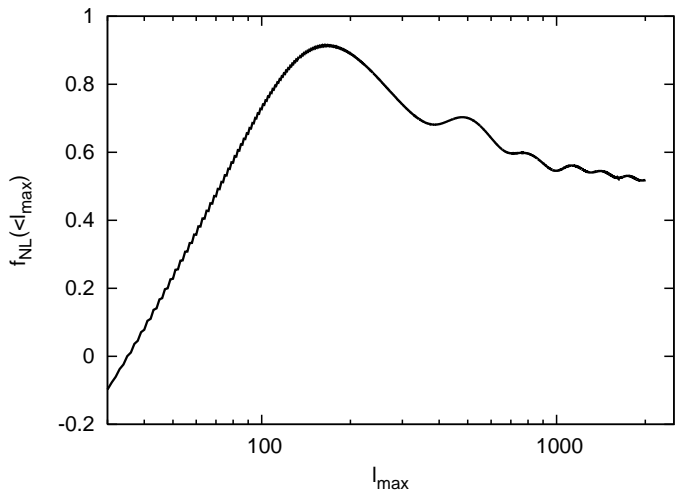}
\end{center}
\caption{\label{fig:fNL}
Contamination of the local primordial bispectrum as measured by
 $f_{NL}^{\rm con}$ (Eq~(\ref{eq:fNL})).
}
\end{figure}

We quantify the contamination of the primordial bispectrum due
to the second-order effects from the products of the first-order
perturbations as follows: we fit the primordial bispectrum template to
the second-order bispectrum, and find the best-fitting $f^{\rm
con}_{NL}$ (``con'' stands for contamination) by
minimizing $\chi^2$ given by
\begin{eqnarray}
\chi^2=\sum_{2\le l_1\le l_2\le l_3}\frac{{\left(f_{NL}B_{l_1l_2l_3}^{prim}-B_{l_1l_2l_3}^{2nd}\right)}^2}{\sigma_{l_1l_2l_3}^2},
\end{eqnarray}
with respect to $f_{NL}$. Here,  $B_{l_1l_2l_3}^{prim}$ is the
local-type primordial bispectrum with $f_{NL}=1$ \cite{ks01}.  We obtain
\begin{eqnarray}
f_{NL}^{\rm con}&=& \frac{1}{N}\sum_{2\le l_1\le l_2\le l_3}\frac{B_{l_1l_2l_3}^{2nd}B_{l_1l_2l_3}^{prim}}{\sigma_{l_1l_2l_3}^2},\nonumber\\
N&=&\sum_{2\le l_1\le l_2\le l_3}\frac{\left({B_{l_1l_2l_3}^{prim}}\right)^2}{\sigma_{l_1l_2l_3}^2}\label{eq:fNL}.
\end{eqnarray}
This is the value of $f_{NL}$ one would find, if one did not know that
the primordial bispectrum did not exist but there was only the
second-order bispectrum from the products of the first-order terms.
In Fig.~\ref{fig:fNL} we show $f_{NL}^{\rm con}$ as a function of the
maximum multipoles, $l_{max}$. We find that $f_{NL}^{\rm con}$ reaches
the maximum value, $\sim 0.9$, when the correlation coefficient reaches
the maximum at $l_{max}\sim 200$, but then decreases to $\sim 0.5$ at
$l_{max}\sim 2000$. Therefore, we conclude that the contamination of the
primordial bispectrum due to the second-order bispectrum is negligible
for CMB experiments.

\begin{figure}[t]
\begin{center}
\includegraphics[width=0.6\textwidth]{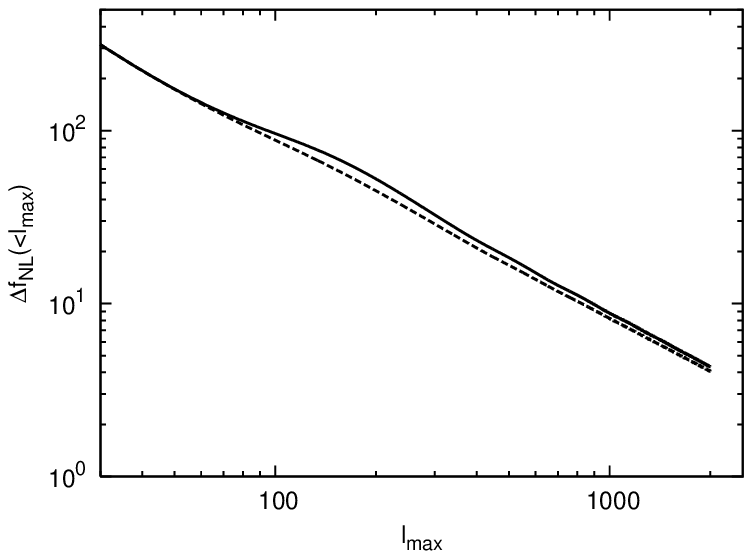}
\end{center}
\caption{\label{fig:DeltafNL}
Projected uncertainty of $f_{NL}$ with (dashed) and without (solid) the
 second-order bispectrum  marginalized over.
}
\end{figure}

Finally, we calculate the 1-$\sigma$ uncertainty of $f_{NL}$, $\Delta
f_{NL}$, with the second-order bispectrum marginalized over. This is
given by $\Delta
f_{NL}=\sqrt{(F^{-1})_{prim,prim}}$. Fig.~\ref{fig:DeltafNL} shows that
an increase in the uncertainty of $f_{NL}$ due to marginalization is
totally negligible. 

\section{Conclusions}
We have presented the general formula of the CMB angular averaged
bispectrum,  Eq.~(\ref{eq:fullbis}), arising from the source terms that
contain second-order perturbations in the Boltzmann equation,
Eq.~(\ref{eq:35}).
In this paper we have considered the source terms that are products of
the first-order perturbations. Since they are products in position
space, similar to the local primordial non-Gaussianity, the predicted
shapes of the angular bispectrum from the products of the first-order
terms are similar to those of the local-type primordial bispectrum, with
cross-correlation coefficients of $\sim 0.5$ and $0.35$ for $l_{max}\sim
200$ and $1000$, respectively.

The predicted signal-to-noise ratio of the products of
the first-order perturbations is small: it reaches only up to $S/N\sim
0.4$ for $l_{max}=2000$, even with an ideal cosmic-variance-limited
experiment. 
The contamination of the local primordial bispectrum is
minimal: the contamination, $f_{NL}^{\rm con}$, is only 
0.9 for $l_{max}=200$ and 0.5 for
$l_{max}=2000$, and an increase in the uncertainty in $f_{NL}$ due to marginalization over the second-order bispectrum is negligible.
This level of the contamination is completely negligible for the present
analysis of the WMAP data \cite{wmap5,smith}.
The contamination is negligible also for the Planck data, for which the
expected 1-$\sigma$ uncertainty is $\Delta f_{NL}\sim 5$, or even for 
the ideal experiment, for which $\Delta f_{NL}\sim 3$ \cite{ks01}.
Therefore, we conclude that the effects of the
products of the first-order perturbations in the Boltzmann equation may
be safely ignored when one tries to extract $f_{NL}$ from the CMB temperature
data. 

We shall present the numerical calculations of the bispectrum that
include the contributions from the intrinsically second-order terms as
well as those from the perturbed recombination, both of which were ignored
in this paper, in future publications.

\ack
D.~N. would like to thank Toshifumi Futamase for helpful
 discussions.  
This work is supported in part by NSF grant PHY-0758153 and the
 Grant-in-Aid for Tohoku University Global Center of Excellence (GCOE)
 Program, ``Weaving Science Web beyond Particle-Matter Hierarchy,'' from
 the Ministry of Education, Culture, Sports, Science and Technology
 (MEXT) of Japan. E.~K. acknowledges support from the Alfred P. Sloan
 Foundation. N.B. and S.M. acknowledge partial financial support by ASI, under contracts I/016/07/0 ``COFIS'' and Planck LFI Activity of Phase E2. 


\section*{Reference}

\begin{thebibliography}{99}

\bibitem{review}
  N. Bartolo, E. Komatsu, S. Matarrese and A. Riotto,
  Phys.\ Rept.\  {\bf 402}, 103 (2004) .

\bibitem{komatsuphd}
  E. Komatsu,
  ph.D. thesis at Tohoku University arXiv:astr-ph/0206039 .

\bibitem{lyth/ungarelli/wands:2003}
  D. H. Lyth, C. Ungarelli and D. Wands,
  Phys.\ Rev.\  D {\bf 67}, 23503 (2003) .

\bibitem{babich/creminelli/zaldarriaga:2004}
  D. Babich, P. Creminelli and M. Zaldarriaga,
  JCAP {\bf 0408}, 009 (2004) .

\bibitem{chen/etal:2007}
  X. Chen, M.-x. Huang, S. Kachru and G. Shiu,
  JCAP {\bf 0701}, 002 (2007) .

\bibitem{holman/tolley:2008}
  R. Holman and A. J. Tolley,
  JCAP {\bf 0805}, 001 (2008) .

\bibitem{maldacena:2003}
  J. M. Maldacena,
  JHEP {\bf 05}, 013 (2003) .

\bibitem{acquaviva/etal:2003}
  V. Acquaviva, N. Bartolo, S. Matarrese and A. Riotto,
  Nucl.\ Phys.\  B {\bf 667}, 119 (2003) .

\bibitem{smith}
  K. Smith, L. Senatore and M. Zaldarriaga,
  arXiv:0901.2572 [astro-ph] .
 
\bibitem{ks01}
  E. Komatsu and D.N.Spergel,
  Phys.\ Rev.\  D {\bf 63}, 063002 (2001) .

\bibitem{lig06}
  M. Liguori, E. Komatsu, S. Matarrese and A. Riotto,
  Phys.\ Rev.\  D {\bf 73}, 043505 (2006) .
  
\bibitem{bar06I}
  N. Bartolo, S. Matarrese and A. Riotto,
  JCAP {\bf 0606}, 024 (2006) .

\bibitem{bar06II}
  N. Bartolo, S. Matarrese and A. Riotto,
  JCAP {\bf 0701}, 019 (2007) .

\bibitem{pitrou070809}
  C. Pitrou, 
  Class. Quant. Grav. {\bf 24}, 6127 (2007); C. Pitrou, Class. Quant. Grav. {\bf 26}, 065006 (2009); C. Pitrou, arXiv:0809.3245 [astro-ph]

\bibitem{bartolo/matarrese/riotto:2006}
  N. Bartolo, S. Matarrese and A. Riotto,
  JCAP {\bf 0605}, 010 (2006) .  

\bibitem{pitrou}
  C. Pitrou, J.-P. Uzan and F. Bernardeau,
  Phys.\ Rev.\  D {\bf 78}, 063526 (2008) .

\bibitem{bartolo}
  N. Bartolo and A. Riotto,
  arXiv:0811.4584 [astro-ph] .

\bibitem{khatri09}
  R. Khatri and B. Wandelt,
  Phys.\ Rev.\  D {\bf 79}, 023501 (2009) .

\bibitem{senatore08I}
  L. Senatore, S. Tassev and M. Zaldarriaga,
  arXiv:0812.3652v1 [astro-ph] .
  
\bibitem{senatore08II}
  L. Senatore, S. Tassev and M. Zaldarriaga,
  arXiv:0812.3658v1 [astro-ph] .

\bibitem{lecture2nd}
  N. Bartolo, S. Matarrese and A. Riotto,
  arXiv:astr-ph/0703496 .
  
\bibitem{cmbfast}
  U. Seljak and M. Zaldarriaga,
  Astrophys.\ J.\  {\bf 469}, 437-444 (1996) .
  
\bibitem{wmap5}
  E. Komatsu {\it et al.},
  Astrophys.\ J.\ S.  {\bf 180}, 330 (2009) .

\end{thebibliography}
  
\end{document}